\documentclass[twocolumn,showpacs,preprintnumbers,amsmath,amssymb,floatfix]{revtex4}

\usepackage{graphicx}
\usepackage{epsfig} 
\usepackage{dcolumn}
\usepackage{bm}
\usepackage{epstopdf}
\usepackage{multirow}
\usepackage{amsmath}
\usepackage{booktabs}

\begin{document}

\preprint{APS/123-QED}

\title{The $C \: ^1 \Sigma ^+$, $A \: ^1 \Sigma ^+$, and $b \: ^3 \Pi_{0^+}$ states of LiRb}

\author{I. C. Stevenson$^1$, D. B. Blasing$^2$, Y. P. Chen$^{1,2,3}$ and D. S. Elliott$^{1,2,3}$}
\affiliation{%
  $^1$School of Electrical and Computer Engineering, $^2$Department of Physics and Astronomy, and $^3$Purdue Quantum Center \\ Purdue University, West Lafayette, IN  47907
}

\date{\today}

\begin{abstract}

We present the first spectroscopic studies of the $C \ ^1\Sigma^+$ electronic state and the $A \ ^1\Sigma^+$ - $b \ ^3\Pi_{0^+}$ complex in $^7$Li - $^{85}$Rb.  Using resonantly-enhanced, two-photon ionization, we observed $v = 7$, 9, 12, 13 and $26-44$ of the $C \ ^1\Sigma^+$ state.  We augment the REMPI data with a form of depletion spectra in regions of dense spectral lines.  The $A \ ^1\Sigma^+$ - $b \ ^3\Pi_{0^+}$ complex was observed with depletion spectroscopy, depleting to vibrational levels $v=0 \rightarrow 29$ of the $A \ ^1\Sigma^+$ state and $v=8 \rightarrow 18$ of the $b \ ^3\Pi_{0^+}$ state.  For all three series, we determine the term energy and vibrational constants.  Finally, we outline several possible future projects based on the data presented here.

\end{abstract}

\maketitle

\section{Introduction}

Ultracold polar molecules have been of  great interest for many years.  They have many applications with quantum logic operations, many-body effects, and collision studies at extremely low energies~\cite{demille2002quantum,ni2010dipolar,krems2009cold}.  Due to the permanent electric dipole moment of these molecules in their rovibrational ground state, the long-range anisotropic interaction between neighbors allows a range of unique physical interactions~\cite{trefzger2009pair,capogrosso2010quantum} that are inaccessible to ultracold atomic systems.  The most recent addition to the bi-alkali family of ultracold molecules is $^7$Li-$^{85}$Rb~\cite{dutta2014formation,altaf2015formation,lorenz2014formation}.  Recent work on this system has revealed a high rate of photoassociation (PA)~\cite{dutta2014formation} and a pathway for direct generation of ground ro-vibronic molecules through PA and spontaneous decay~\cite{stevenson2016direct}.  Much of this work makes use of a resonantly coupled $2(1) - 4(1)$ state for PA~\cite{stevenson2016direct}.  We have also observed short-range PA through the $d \ ^3\Pi_{\Omega}$ states, leading to population in the low-lying vibrational levels of the metastable $a \ ^3\Sigma^+$ state~\cite{blasingsce16}.

In order to achieve many of the goals for ultracold polar molecules, it is essential to generate and confine a large number of molecules.  For generation, trapping, and control of the molecules, we must explore the various electronic potentials, with special interest in singlet-triplet mixed states that can be useful in transferring population from triplet to singlet states (or vice versa).  In the present work, we explore the $C \ ^1 \Sigma ^+$, $A \ ^1 \Sigma ^+$, and $b \ ^3 \Pi_{0^+}$ states, all previously unexplored in LiRb.  For each of these potentials we determine the term energies and vibrational constants.  With the exploration of the levels discussed in the present work, only one low-lying electronic state has not yet been studied, the $c \: ^3 \Sigma^+$ state.  In Table~\ref{tab:StatesObserved} we list the eleven lowest energy states of the LiRb system and the vibrational levels that have been observed for each. 
\begin{table}
	\centering

	\begin{tabular}{c c c c}
		\hline \hline
		Electronic & Vib. levels & & Atomic   \\
		 state & observed & Refs. & asymptote\\ \midrule[1.5pt]
		$X \: ^1 \Sigma^+$ & 0-45 &   \cite{ivanova2011x,dutta2011laser} &  Li $2s \ ^2S + $  \\ 
		   $a \: ^3 \Sigma^+$ & 2-13$^{\ast}$ & \cite{altaf2015formation,blasingsce16,ivanova2011x} & \hspace{0.2in} Rb $5s \ ^2S  $  \\
	   \hline
	   
	   $b \: ^3 \Pi$ & 9-19 &  this work  & \\
	    	$A \: ^1 \Sigma^+$ & 0-29 & this work &   Li $2s \ ^2S + $  \\
		
	 $c \: ^3 \Sigma^+$ & - &  - & \hspace{0.2in} Rb $5p \ ^2P  $  \\	
		$B \: ^1 \Pi$ & 0-22 & \cite{ivanova2013b1pi,dutta2011laser,lorenz2014formation} & \\
		\hline
			   
		$C \: ^1 \Sigma^+$ & 7-13$^{\ast}$, 26-44 & this work & \\
		  $D \: ^1 \Pi$ & 0-15 & \cite{ivanova2013b1pi,stevenson2016d,lorenz2014formation}  &  Li $2p \ ^2P + $  \\
		
		 $d \: ^3 \Pi$ & 0-22 &  \cite{stevenson2016d} & \hspace{0.2in} Rb $5s \ ^2S  $  \\
		 \hline

	 $f \: ^3 \Pi$ & 0-10  & \cite{altaf2015formation} & Li $2s \ ^2S + $\\
		
		 $g \: ^3 \Sigma^+$ & 0-5 & \cite{altaf2015formation} & \hspace{0.2in} Rb $4d \ ^2D  $ \\
				
 \hline \hline
	\end{tabular}
	\caption{List of low-lying states of LiRb that have been observed experimentally.  Ranges of vibrational quantum numbers marked with an asterisk are not inclusive. Weakly-bound vibrational levels of many of these potentials used for photoassociation of ultracold LiRb molecules are not included.}
	\label{tab:StatesObserved}
\end{table} 

In the past, our spectroscopic studies of LiRb~\cite{dutta2011laser,altaf2015formation,stevenson2016d,stevenson2016direct} have been driven by our goal of producing large numbers of ultracold molecules in the ro-vibronic ground state.  Our work here is no different; we evaluate possible stimulated Raman adiabatic passage (STIRAP) schemes through the $C \ ^1\Sigma^+$ spectra for the transfer of population from loosely bound singlet molecules, produced at a high rate via photoassociation~\cite{dutta2014photoassociation,stevenson2016direct}, to low vibrational levels of the $X \ ^1\Sigma^+$ state.  For $v$ between 20 and 25 of the $C \ ^1\Sigma^+$ state, the calculated Franck-Condon factors (FCFs) for both legs of this Raman transition are favorable.  This forms an ideal transfer pathway in LiRb.  We also evaluate various vibrational levels of the $A \ ^1\Sigma^+$ state, which have played a key role in our previous determination~\cite{stevenson2016direct} of the vibrational distribution of population in $X \ ^1\Sigma^+$ ground state.  As part of this study, we find and evaluate strong spin-orbit coupling for certain vibrational levels of the $A \ ^1\Sigma^+$ - $b \ ^3\Pi_{0^+}$ complex.  Finally, the present measurements of higher vibrational levels of the  $b \ ^3\Pi_{0^+}$ state narrow a future search for the lowest several vibrational levels.  Two possible applications for these mixed states are: a direct molecular laser cooling scheme proposed by You \textit{et al.}~\cite{you2016ab} and short range photoassociation to produce ground state molecules.

The structure of this paper is as follows.  In Sec.~\ref{sec:ExpParams} we provide the relevant details of our experimental setup. In Sec.~\ref{sec:CState}, we present and discuss our data on the $C \ ^1\Sigma^+$ electronic state.  In Sec.~\ref{sec:Astate}, we present our data on the $A \ ^1\Sigma^+$ and $b ^3 \Pi_{0^+}$ states.  In Sec.~\ref{sec:Abcomplex}, we present our findings on $A \ ^1\Sigma^+$ - $b \ ^3\Pi_{0^+}$ mixing.  Finally in Sec.~\ref{sec:Outlook}, we conclude and provide an outlook for future experiments in LiRb.

\begin{figure}[t]
	\includegraphics[width=8.6cm]{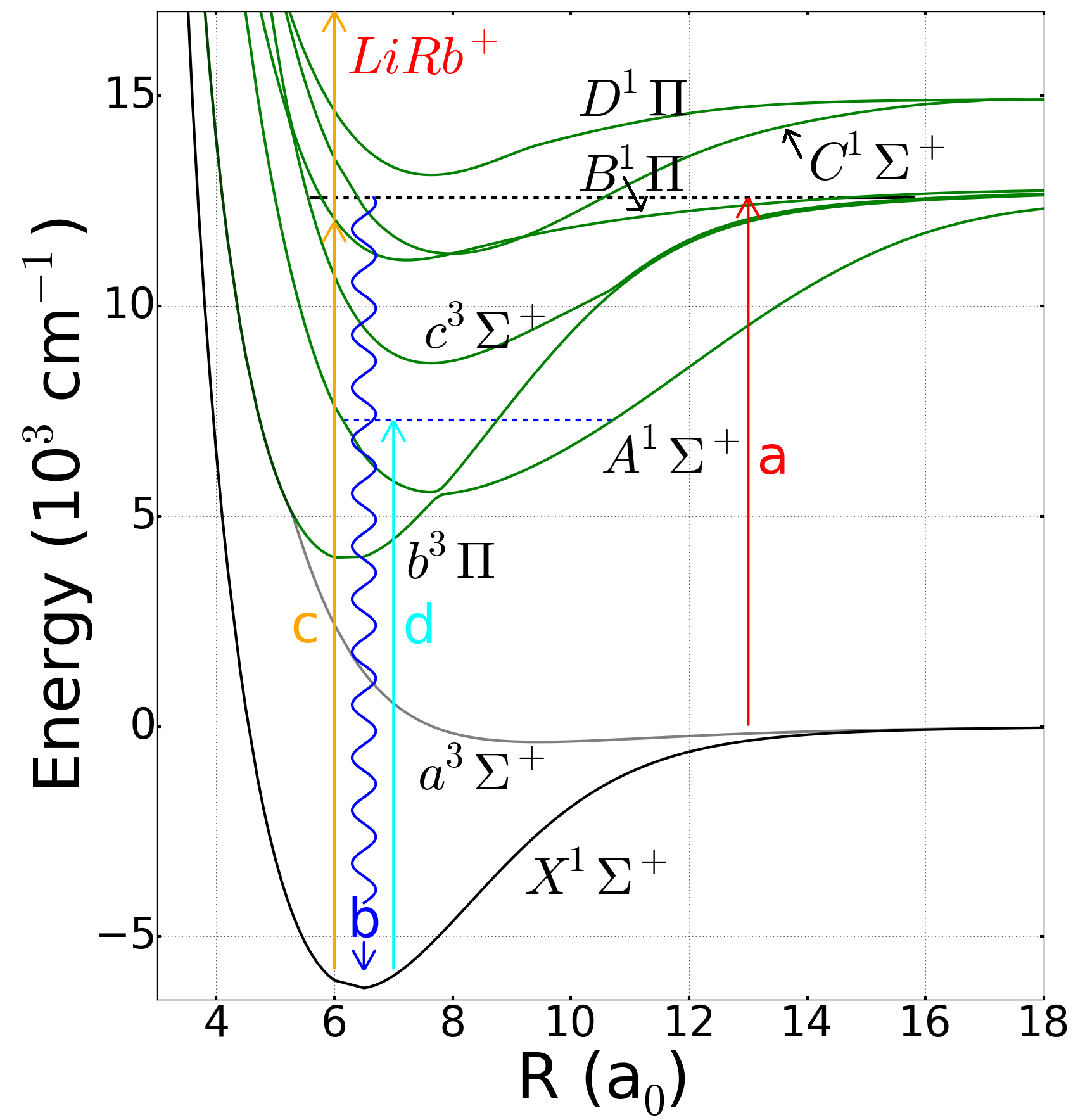}\\
	\caption{(Color on-line) Energy level diagram for the LiRb molecule showing relevant PECs from Ref.~\cite{korek2009theoretical}.  Vertical lines show various optical transitions, including {\bf (a)} photoassociation of molecules below the  Rb D$_1$ asymptote; {\bf (b)} spontaneous decay of excited state molecules leading to a wide range of $X \: ^1 \Sigma ^+$ vibrational levels, drawn for $v^{\prime \prime}=2$; {\bf (c)} REMPI to ionize LiRb molecules, labeled $\nu_{c}$ in other figures; and {\bf (d)} excitation to deplete the REMPI signal.  The black dashed line represents the 2(1) - 4(1) PA resonance, while the blue dashed line represents the depletion state.}
	\label{fig:PEC}
\end{figure}

\section{Experiments} \label{sec:ExpParams}

We work out of a dual-species MOT, trapping $\sim 5 \times 10^7$ Li atoms and $\sim 2 \times 10^8$ Rb atoms in a cloud $\lesssim$1 mK in temperature and 1 mm in diameter~\cite{altaf2015formation}.  Our Rb MOT is a spatial dark spot MOT~\cite{ketterle1993high}.  We photoassociate Li and Rb atoms to form LiRb molecules using a 150 mW external-cavity diode laser (ECDL), which we tune to the $4(1) \ v=-16 \ J=1$ mixed state at $\nu_a = 12574.85$ cm$^{-1}$~\cite{stevenson2016direct}.  Spontaneous decay from the PA state forms bound LiRb molecules in many vibrational levels of the $X \ ^1\Sigma^+$ ground state.  We use $v$ and $J$ (without a prime) to denote the vibrational and rotational levels of the PA resonances (and for these vibrational numbers, we count down from the asymptote using negative integers), $v^{\prime}$ and $J^{\prime}$ to denote vibrational and rotational labeling of other excited electronic states, and $v^{\prime \prime}$ and $J^{\prime \prime}$ to denote the vibrational and rotational levels of the $X \: ^1 \Sigma ^+$ state.  

We use resonantly-enhanced-multiphoton-ionization to ionize these molecules, which we detect with a time-of-flight spectrometer and microchannel plate detector.  The laser used to drive the ionization process is a Nd:YAG pumped, pulsed dye laser.  This laser is tunable in the wavelength range between 550 - 583 nm when loaded with an R590 dye ($\nu_c = 18150$ - 17150 cm$^{-1}$), or between 662 - 709 nm ($\nu_c = 15100$ - 14100 cm$^{-1}$) when loaded with an LDS 698 dye.  The repetition rate of this laser is 10 Hz, and it delivers $\sim$3 mJ/pulse to the MOT region in a 4 mm diameter beam.  When the frequency of this laser is one-photon resonant with a transition from a state that is populated by spontaneous decay from the PA state to an intermediate bound state (usually one of the vibrational levels of the $C \: ^1 \Sigma^+$, $B \: ^1 \Pi$ or $D \: ^1 \Pi$ electronic potentials), absorption of two photons ionizes the molecule.  The second step in this ionization process can be driven by the dye laser pulse (that is, one-color ionization, which we designate REMPI), or by the 532 nm pump laser pulse (two-color ionization, which we call RE2PI).  In the latter case, we decrease the dye laser pulse energy to $\sim$0.5 mJ, and adjust the timing between the two pulses to optimize the ionization signal.  The frequency resolution of REMPI or RE2PI features is $\sim$0.5 cm$^{-1}$, limited by the linewidth of the pulsed dye laser. 

\begin{figure*}[t]
	\includegraphics[width=\textwidth]{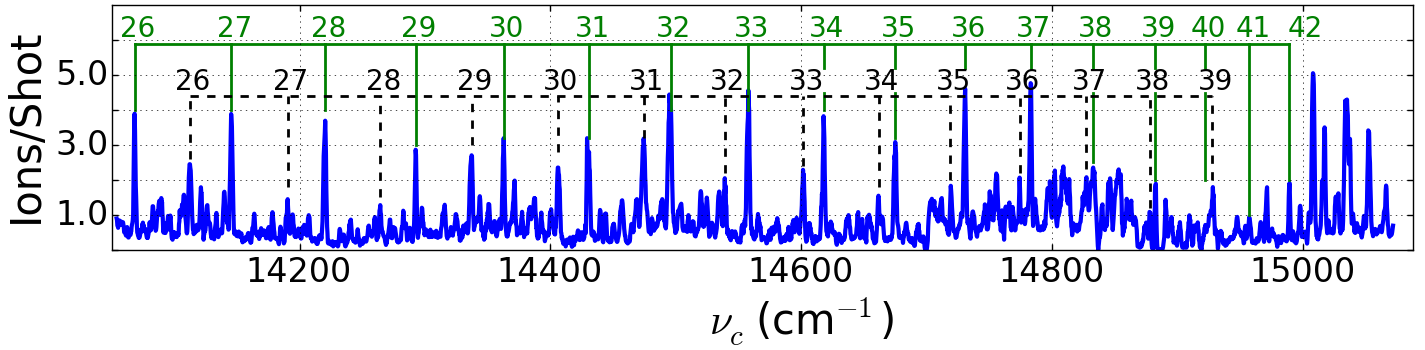} \\ 
	\caption{(Color on-line) RE2PI spectra from $v^{\prime \prime}=42$ and 43 to the $C \ ^1\Sigma^+$ electronic state. Green solid lines and numbers label transitions $C \ ^1\Sigma^+ \ v^{\prime} \leftarrow v^{\prime \prime}=43$;  black dashed lines and numbers label transitions $C \ ^1\Sigma^+ \ v^{\prime} \leftarrow v^{\prime \prime}=42$.  }
	\label{fig:v43progression}
\end{figure*}

We also use a form of depletion spectroscopy to augment our RE2PI studies.  In this technique, we add a cw narrow-band depletion laser to the configuration.  As previously demonstrated in Refs.~\cite{deiglmayr2010permanent,stevenson2016direct}, the depletion laser reduces the REMPI ion count by resonantly transferring population from the active initial state in REMPI to other bound states.  Our depletion laser is a 300 mW cw Ti:Sapphire laser that we tune between 740 - 905 nm.  We co-propagate the PA and depletion lasers, and then focus them to a spot size of about 200 $\mu$m in diameter in the center of the MOTs.  Depletion spectroscopy has a high spectral resolution, limited primarily by the natural lifetime or power-broadened linewidth of the molecular transition.

The population distribution of the initial states limits excited states accessible with the depletion laser.  Since the ground state molecules are mostly in the $J^{\prime \prime}=0$ rotational state~\cite{stevenson2016direct}, selection rules allow optical transitions to $J^{\prime}=1$ rotational states.  Transitions from $J^{\prime \prime}=2 \rightarrow J^{\prime}=$1 or 3 are barely detectable above the shot noise level, so we are unable to determine the rotational constants of the excited state. Additionally we are limited by the spin of our molecules starting in the $X \ ^1\Sigma^+$ state.  For transitions to $b \ ^3\Pi_{\Omega}$, where $\Omega = 0^+$, $0^-$, 1, and 2, only $\Omega = 0^+$ peaks appear in the spectra, borrowing strength from the nearby $A \ ^1\Sigma^+$ vibrational levels.  $\Omega$ is the projection of the total electronic angular momentum (orbital plus spin) onto the internuclear axis.  Transitions to $\Omega = 0^-$ or 2 are not allowed.  Transitions to $\Omega = 1$ are possible due to second-order spin-orbit mixing with mixed $A \ ^1\Sigma^+ - b \ ^3\Pi_{0^+}$ states, but they are too weak to observe.  

Besides shot noise in the ion count, two sources of uncertainty in depletion peak positions deserve mention: frequency drifts of both the PA laser and the REMPI laser.  The PA laser frequency was maintained with an electronic lock.  Drift in the REMPI laser frequency was small, slow and nearly linear, so we were able to compensate in post-processing.  Therefore, the primary limitation to the precision of these measurements was the shot noise in the signal.

\section{$C \ ^1\Sigma^+$ discussion}\label{sec:CState}
We make use of the population in the $v^{\prime \prime}=42$ and 43 levels of the $X \ ^1\Sigma^+$ state, formed by spontaneous decay after the PA step, to record the positions of the $v^\prime=26-45$ vibrational lines of the $C \ ^1\Sigma^+$ state.  To detect these molecules we use RE2PI with the LDS 698 dye in the dye laser, which ionizes the molecules through the loosely bound vibrational levels of the $C \ ^1\Sigma^+$ electronic state.  Thus we are able to measure the energies of the vibrational levels of the $C \ ^1\Sigma^+$ state relative to $v^{\prime \prime}=42$ and 43.

As part of our work here, we measured the binding energies of $v^{\prime \prime}=42$ and $v^{\prime \prime}=43$ with the highest precision to date.  In Ref.~\cite{stevenson2016direct}, we measured the binding energy of the $4(1) \ v=-16 \ J=-1$ PA resonance to be 12574.85 (0.02) cm$^{-1}$, and in this project we measured the frequency of the depletion transition $B \ ^1\Pi \ v^{\prime}=20 \leftarrow X \ ^1\Sigma^+ \ v^{\prime \prime}=43$ to be 12711.71 (0.02) cm$^{-1}$; the difference between these two frequencies gives us the binding energy, $\nu_{43}=-136.86 \ (0.02)$ cm$^{-1}$ ($4(1) \ v=-16 \ J=-1$ is the Hund's case (c) labeling of the $B \ ^1\Pi \ v^{\prime}=20$ state).  Additionally using our RE2PI spectra in Fig.~\ref{fig:v43progression}, we extracted the energy difference between $v^{\prime \prime}=42$ and $v^{\prime \prime}=43$ to be 44.1 (0.1) cm$^{-1}$, which implies the binding energy of $\nu_{42}=-181.0 \ (0.1)$ cm$^{-1}$.  This is in good agreement, but of higher precision, to the prior measurements by Ref.~\cite{dutta2011laser,ivanova2011x} which measured a binding energy of $\nu_{43}=-137 \ (4)$ cm$^{-1}$ and the energy difference between $v^{\prime \prime}=42$ and $v^{\prime \prime}=43$ to be 44.04 cm$^{-1}$. 

\begin{figure}[t]
	\includegraphics[width=8.6cm]{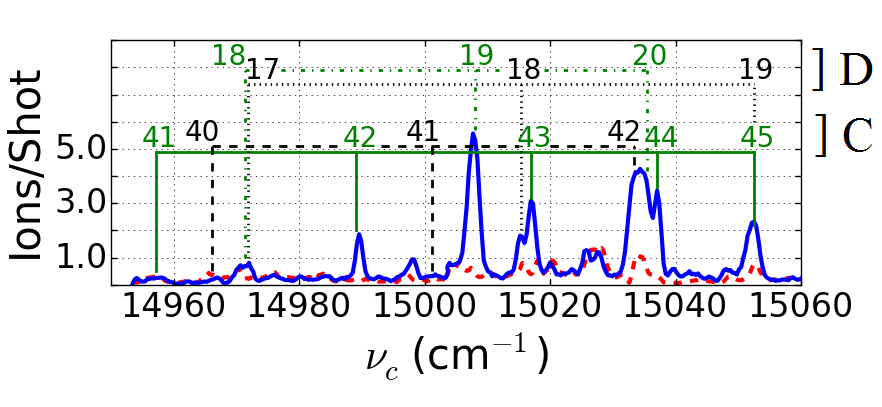} \\ 
	\caption{(Color on-line) Depletion spectroscopy of RE2PI data on transitions from $v^{\prime \prime}=42$ and 43 to the $C \ ^1\Sigma^+$ and $D \ ^1\Pi$ electronic states.  For these data, the blue solid trace is the original data from Fig.~\ref{fig:v43progression}, while the red dashed trace is the RE2PI data retaken in the presence of a depletion laser driving the $B \ ^1\Pi \ v^{\prime}=20 \leftarrow v^{\prime \prime}=43$ transition.  Reduction of a peak height in the presence of the depletion laser indicates that the initial state is $v^{\prime \prime}=43$.  Assignments of these lines are labeled as follows: green, solid lines, $C \ ^1\Sigma^+ \leftarrow v^{\prime \prime}=43$; black, dashed lines, $C \ ^1\Sigma^+ \leftarrow v^{\prime \prime}=42$; 
		black, dotted lines, $D \ ^1\Pi \leftarrow v^{\prime \prime}=42$; green, dot-dashed lines, $D \ ^1\Pi \leftarrow v^{\prime \prime}=43$. }
	\label{fig:v43asymptote}
\end{figure}

We show the RE2PI spectrum of $v^\prime=26$ - 42 of the $C \ ^1\Sigma^+$ state in Fig.~\ref{fig:v43progression}.  Two progressions dominate, one from $v^{\prime \prime}=42$ and the other from $v^{\prime \prime}=43$.  They are spaced by 44.1 cm$^{-1}$, precisely the energy difference between $v^{\prime \prime}=42$ and 43~\cite{dutta2011laser}.  We have marked these series in Fig.~\ref{fig:v43progression} with black dashed lines and green solid lines, respectively.  The integers indicate our assignments of the intermediate RE2PI state, which is a vibrational level $v^{\prime}$ of the $C \ ^1\Sigma^+$ state.  The weak, unlabeled RE2PI lines in this spectrum originate from $v^{\prime \prime}=38-41$, which are also populated by the PA resonance we used.  These qualitative features are consistent with the calculation of the FCF for the spontaneous decay from the PA state to $X \ ^1\Sigma^+$, which yields values for $v^{\prime \prime}=43$ and 42 (0.35 and 0.13, respectively) much greater than for any other vibrational level~\cite{stevenson2016direct}.  For frequencies significantly below the Rb 5S + Li 2P asymptote, the line density of the spectrum in Fig.~\ref{fig:v43progression} is very low allowing us to measure the energies of $v^{\prime}=26 - 40$ twice, thus resulting in $0.5/\sqrt{2} = 0.3$ cm$^{-1}$ precision.  

As the $v^{\prime \prime}=43$ series converges on the Rb 5S + Li 2P asymptote, line congestion increases.  This occurs near $\nu_c = \nu_{D_1} + \nu_{43}  \simeq 15050$ cm$^{-1}$, where $\nu_{D_1} = 14904$ cm$^{-1}$ is the frequency of the atomic Li $D_1$ line.  To assign these peaks we used depletion spectroscopy to identify sets of RE2PI peaks that share a common initial state, as shown in Fig.~\ref{fig:v43asymptote}.  For these data, the blue solid trace is the original data from Fig.~\ref{fig:v43progression}, while the red dashed trace is the RE2PI data retaken in the presence of a depletion laser driving the $B \ ^1\Pi \ v^{\prime}=20 \leftarrow v^{\prime \prime}=43$ transition.  The peaks originating from $v^{\prime \prime}=43$ (marked by the green solid and green dot-dashed lines in Fig.~\ref{fig:v43asymptote}) largely vanish upon introduction of the depletion laser.  Careful inspection of Fig.~\ref{fig:v43asymptote} shows that all the strong peaks disappear after we introduce the depletion laser; this is consistent with the picture told by FCFs, that is there is three times more population in $v^{\prime \prime}=43$ than in $v^{\prime \prime}=42$.  The peak for the $C \ ^1\Sigma^+ \ v^{\prime}=45 \leftarrow v^{\prime \prime}=43$ is tentative, since the energy of this peak is above the asymptote.  Still, we note that the peak is strongly depleted by the depletion laser, the spacing is about right for this progression, and the long range-potential for the $C \ ^1\Sigma^+$ PEC is expected to be repulsive~\cite{korek2009theoretical}, that is $C_6 >0$, which could lead to a quasi-bound state above the asymptote.  The $D \ ^1\Pi$ potential from Ref.~\cite{ivanova2013b1pi} provides a good guide in this region, with an average uncertainty of 2.5 cm$^{-1}$, which allows us to identify candidate lines for transitions to the $C \ ^1\Sigma^+$ state.  The unassigned lines in Fig.~\ref{fig:v43asymptote} most likely result from transitions $D \ ^1\Pi \ v^{\prime}=18 - 20 \leftarrow v^{\prime \prime}=38 - 41$; these initial states are weakly populated by the PA resonance, and the calculated FCFs for these lines are strong.  

For $v^{\prime}<13$, we used the R590 dye in the pulsed dye laser, and excited from the $v^{\prime \prime} $ = 2 or 3 level of the ground state.  Otherwise, the measurements were similar to those described above.  The density of lines in these spectra was great enough to require depletion spectroscopy to make line-assignments with confidence.  Table~\ref{tab:C1Sigma+} contains a summary of our observed energies of the $v^{\prime}=7, \ 9, \ 12, \ 13$ and $26-45$ vibrational levels of the $C \ ^1\Sigma^+$ state.

\begin{table} [t]
	\centering
	\begin{tabular}{ccc}
		\hline \hline
		$v^{\prime}$ & $T_v$ (cm$^{-1}$) & $\Delta$E (cm$^{-1}$) \\ \midrule[1.5pt]
		7 & 12131.2 &  \\ 
		9 & 12344.2 &  \\
		12 & 12655.8 & 102.4 \\
		13 & 12758.2 &  \\
		&& \\
		26 & 13931.3 & 77.6 \\
		27 & 14008.9 & 74.4 \\
		28 & 14083.2 & 72.5 \\ 
		29 & 14155.7 & 69.5 \\
		30 & 14225.3 & 68.3 \\
		31 & 14293.6 & 64.8 \\
		32 & 14358.4 & 62.3 \\
		33 & 14420.6 & 60.0 \\
		34 & 14480.6 & 57.3 \\
		35 & 14537.9 & 55.5 \\
		36 & 14593.4 & 52.6 \\
		37 & 14646.0 & 50.0 \\
		38 & 14696.0 & 49.9 \\
		39 & 14745.9 & 44.9 \\
		40 & 14784.9 & 38.9 \\
		41 & 14820.1 & 32.0 \\
		42 & 14852.1 & 27.8 \\
		43 & 14879.9 & 20.2 \\
		44 & 14900.1 & 15.4\\ 
		45 & 14915.5 &  \\ \hline \hline
	\end{tabular}
	\caption{Measured vibrational energies of the $C \ ^1\Sigma^+$ electronic state.  The uncertainty in $T_v$ is 0.5 cm$^{-1}$ for $v^{\prime}=7-13$ and $40-45$ and is 0.3 cm$^{-1}$ for $v^{\prime}=26-40$.  The vibrational numbering was chosen to smoothly connect to the deeply bound vibrational levels measured in Ref.~\cite{ivanova2013b1pi}.  Energies are referenced to the Rb 5S + Li 2S atomic asymptote.}
	\label{tab:C1Sigma+}
\end{table} 

There have been no previous direct observations of the $C \ ^1\Sigma^+ \ v^{\prime}$ states.  In Ref.~\cite{ivanova2013b1pi}, the authors observed perturbations to the rotational lines of the $B \ ^1\Pi \leftarrow X \ ^1\Sigma^+$ spectrum, from which they extracted the binding energies of $C \ ^1\Sigma^+ \ v^{\prime} \simeq 0 - 13$, and calculated a $C \ ^1\Sigma^+$ PEC.  We see good agreement with the PEC of Ref.~\cite{ivanova2013b1pi} for $v^{\prime}<13$.  We fitted the $C \ ^1\Sigma^+ \ v^{\prime} = 7-40$ line positions listed in Table~\ref{tab:C1Sigma+} to  
\begin{eqnarray}\label{eq:termenergy}
   T(v) & = & T_e + \omega_e (v + 1/2) \\ 
     & & \hspace{0.2in} - \omega_e x_e (v + 1/2)^2 + \omega_e y_e (v + 1/2)^3, \nonumber
\end{eqnarray}
and extracted both the term energy $T_e$ and the vibrational constants, and made a smooth connection across the gap from $v^{\prime}=13-26$, which lead to the vibrational numbering we report.  The energies of states $v^{\prime}=41-45$ do not fit Eq.~(\ref{eq:termenergy}) well and were omitted from the fit.  We present the term energy and molecular constants in Table~\ref{tab:molecularConstants}.
\begin{table*}[t]
	\centering
	\begin{tabular}{ccccccccc}
		\hline \hline
		& \multicolumn{3}{c}{$C \ ^1\Sigma^+$} & \multicolumn{2}{c}{$A \ ^1\Sigma^+$} & \multicolumn{3}{c}{$b \ ^3\Pi_{0^+}$} \\
		& Exp. & Exp.~\cite{ivanova2013b1pi} &  Th.~\cite{korek2009theoretical} & Exp. & Th.~\cite{korek2009theoretical} & Exp. & Th.~\cite{korek2009theoretical} & Th.~\cite{you2016ab} \\ \midrule[1.5pt]
		$\omega_e$ (cm$^{-1}$) & 115.4 (0.9) & 113.8 (0.2) & 113.4 & 117.3 (0.6) & 117.9 & 195.1 (1.0) & 190.6 & 188.3 \\
		$x_e \times 10^3$ & 3.1 (0.4) & & 2.3 & 3.1 (0.4) & 3.3 & 4.3 (0.2) & 2.9 & 3.4 \\
		$y_e \times 10^6$ & -78 (5) & & -87 & 15 (9) & 23 &  & -18.4 & -29.7 \\
		$T_e$ (cm$^{-1}$) & 11288 (13) & 11302 (4) & 11237.4 & 5756.6 (2.2) & 5537.7 & 4083 (7) & 3962.3 & 4180.4 \\ \hline \hline
	\end{tabular}
	\caption{Molecular constants fitted to our data, compared to fits of other experiments~\cite{ivanova2013b1pi} or to theoretical predictions from \textit{ab initio} calculations from Refs.~\cite{you2016ab} and~\cite{korek2009theoretical}.  The state energies are given by Eq.~(\ref{eq:termenergy}).  For fitting $C \ ^1\Sigma^+$ we omitted $v^{\prime}=41 - 45$ because the energies of these states fall outside the range over which Eq.~(\ref{eq:termenergy}) is valid.  When fitting $b \ ^3\Pi_{0^+}$, the vibrational numbering is uncertain since we have not observed $v^{\prime}<8$.  We chose vibrational numbering to force a well depth between the two \textit{ab initio} predictions.  Additionally, we omitted $v^{\prime}=10$, 12, 14 and 16 from the fit because these levels are strongly perturbed, as shown in Fig.~\ref{fig:vibSpacing}(b).  For these data, a value of $y_e$ is not significant and not included in the table.} 
	\label{tab:molecularConstants}
\end{table*}

\section{${\bf A \: ^1\Sigma^+}$,  ${\bf b \: ^3\Pi_{0^+}}$ Discussion} \label{sec:Astate}

In this section, we discuss a set of measurements in which we use depletion spectra to identify new excited states.  This is in contrast to the previous section where we used depletion spectra only to identify common initial states.  In total we deplete two REMPI transitions exciting deeply bound $X \ ^1\Sigma^+$ ground states to the $A \: ^1\Sigma^+ \ v^{\prime}=0-29$ and the $b \: ^3\Pi_{0^+} \ v^{\prime}=8-18$ levels.  For $A \: ^1\Sigma^+ \ v^{\prime}=19-29$ states, we used $v^{\prime \prime}=10$ as the initial state for depletion, and tuned the REMPI laser to the $D \ ^1\Pi \ v^{\prime}=4 \leftarrow v^{\prime \prime}=10$ transition.  For $A \: ^1\Sigma^+ \ v^{\prime}=0-18$ and $b \: ^3\Pi_{0^+} \ v^{\prime}=8-18$ states, we used $v^{\prime \prime}=2$ as the initial state, and tuned the REMPI laser to the $B \ ^1\Pi \ v^{\prime}=14 \leftarrow v^{\prime \prime}=2$ transition.  We were unable to search for $v^{\prime}<8$ of the $b \: ^3\Pi_{0^+} $, which fall outside the tuning range of our Ti:Sapphire laser.  The $A \: ^1\Sigma^+$ and $b \: ^3\Pi_{0^+}$ series are readily distinguishable since the $A \: ^1\Sigma^+$ lines are extremely power broadened.  In fact, their width greatly facilitated their search, with guidance from the \textit{ab initio} calculations of Ref.~\cite{korek2009theoretical}.  Such broad linewidths of the peaks allowed us to increment the depletion laser frequency in 10 GHz steps when searching for resonances.  Then, we made low intensity measurements of the line positions to remove any line shifts caused by saturation effects.  We tabulate the energies of the $A \: ^1\Sigma^+ \ v^{\prime} \ J^{\prime} = 1$ states in Table~\ref{tab:masterAssignments}, and the energies of the $b ^3\Pi_{0^+} \ v^{\prime} \ J^{\prime} = 1$ states in Table~\ref{tab:b3PiAssignments}.  

\begin{table}[b]
	\centering
	\begin{tabular}{ccc}
		\hline \hline
		$v^{\prime}$ & $T_v$ + 2 $B_v$ (cm$^{-1}$) & $\Delta$E (cm$^{-1}$)\\ \midrule[1.5pt]
		0 & 5808.10 & 122.52 \\
		1 & 5930.62 & 127.55 \\
		2 & 6058.17 & 104.47 \\
		3 & 6162.64 & 117.18 \\
		4 & 6279.82 & 106.67 \\
		5 & 6386.50 & 118.75 \\
		6 & 6505.25 & 111.78 \\
		7 & 6617.02 & 110.61 \\
		8 & 6727.63 & 111.01 \\
		9 & 6838.64 & 110.88 \\
		10 & 6949.52 & 106.71 \\
		11 & 7056.23 & 111.41 \\
		12 & 7167.64 & 110.58 \\
		13 & 7278.21 & 104.49 \\
		14 & 7382.65 & 110.04 \\
		15 & 7492.70 & 108.41 \\
		16 & 7601.10 & 104.31 \\
		17 & 7705.41 & 107.04 \\
		18 & 7812.45 & 105.84 \\
		19 & 7918.29 & 104.34 \\
		20 & 8022.63 & 104.44 \\
		21 & 8127.07 & 103.83 \\
		22 & 8230.90 & 103.14 \\
		23 & 8334.04 & 102.90 \\
		24 & 8436.95 & 102.84 \\
		25 & 8539.78 & 100.67 \\
		26 & 8640.45 & 101.80 \\
		27 & 8742.26 & 101.80 \\
		28 & 8844.06 &  97.97 \\
		29 & 8942.03 &  \\ \hline \hline
	\end{tabular}
	\caption{Experimental assignments for the location of $A \ ^1\Sigma^+ \ v^{\prime} \ J^{\prime}=1$ based on our depletion data.  The uncertainty for all assignments is 0.02 cm$^{-1}$.  Energies are referenced to the Rb 5S + Li 2S atomic asymptote.}
	\label{tab:masterAssignments}
\end{table}
\begin{table}[b]
	\centering
	\begin{tabular}{ccc}
		\hline \hline
		$v^{\prime}$ & $T_v$ + 2 $B_v$ (cm$^{-1}$) & $\Delta$E (cm$^{-1}$) \\ \midrule[1.5pt]
		8 & 5682.01 & 179.41 \\ 
		9 & 5861.42 & 170.83 \\
		10 & 6032.25 & 185.39 \\
		11 & 6217.65 & 180.22 \\
		12 & 6397.87 & 168.08 \\
		13 & 6565.95 & 172.92 \\
		14 & 6738.87 & 169.28 \\
		15 & 6908.16 & 170.78 \\
		16 & 7078.94 & 162.65 \\
		17 & 7241.59 & 167.68 \\
		18 & 7409.27 &  \\ \hline \hline
		
	\end{tabular}
	\caption{Experimental assignments for the location of $b \ ^3\Pi_{0^+} \ v^{\prime} \ J^{\prime}=1$ based on our depletion data.  Uncertainty for assignments is 0.02 cm$^{-1}$, vibrational designation is approximate.  Energies are referenced to the Rb 5S + Li 2S atomic asymptote.}
	\label{tab:b3PiAssignments}
\end{table}

\begin{figure}[t]
	\includegraphics[width=8.6cm]{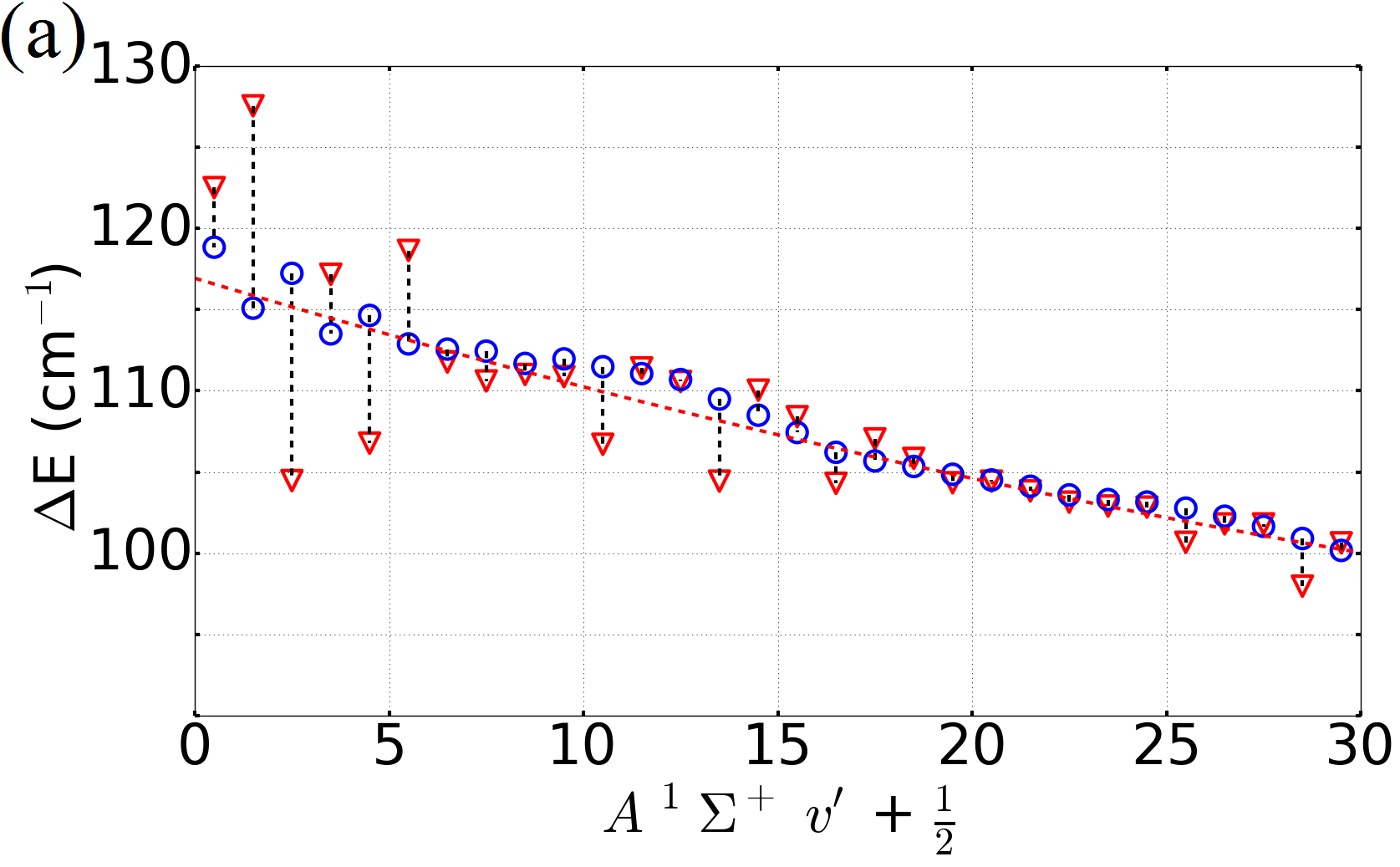}\\
	\includegraphics[width=8.6cm]{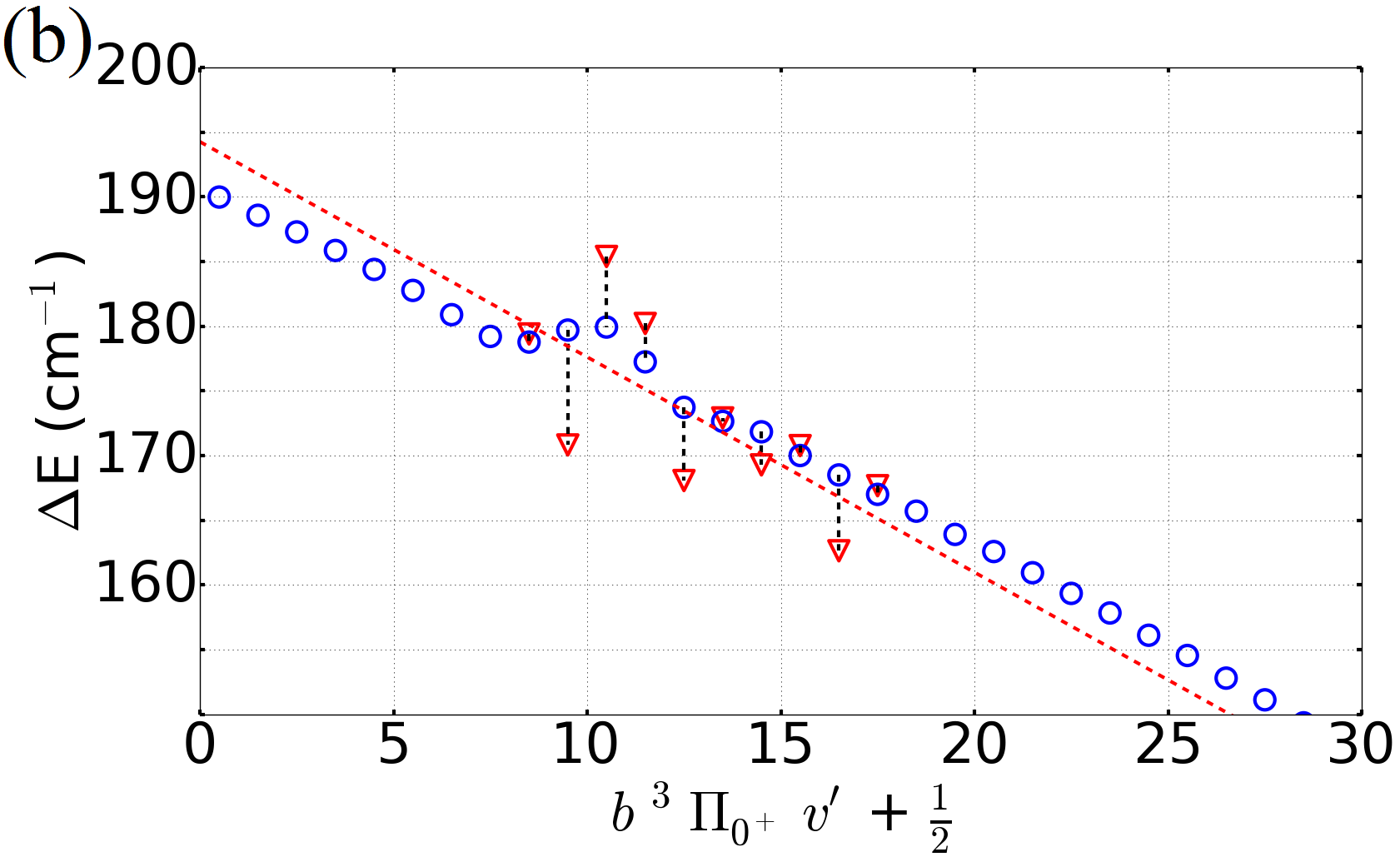}
	\caption{(Color on-line) Comparison of theoretical vibrational spacings (blue circles) to our measurements (red triangles) for (a) $A \ ^1\Sigma^+$ and (b) $b \ ^3\Pi_{0^+}$.  Black dashed vertical lines connect datasets to guide the eye.  The dashed red line is the fit using Eq.~(\ref{eq:termenergy}) and the molecular constants shown in Table~\ref{tab:molecularConstants}.  Large deviations in experiment from the theory are driven by perturbations.  The unevenness of the theory predictions in the $v^{\prime}=0-4$ region for $A \ ^1\Sigma^+$ and $v^{\prime}=7-12$ region for $b \ ^3\Pi_{0^+}$ is an artifact caused by imperfect conversion from adiabatic to diabatic PECs.  Finally, we note that (b) suggests that our extracted constants, $\omega_e$ and $x_e$, for the $b \ ^3\Pi_{0^+}$ state are too large.  Our data for this state is limited, and it appears that if both were smaller, the fit would better match the theory (blue circles).}
	\label{fig:vibSpacing}
\end{figure}

In Fig.~\ref{fig:vibSpacing}, we show the experimental and theoretical~\cite{korek2009theoretical} energy spacing between the vibrational lines of (a) the $A \: ^1\Sigma^+$ series and (b) the $b \ ^3\Pi_{0^+}$ series.  In this figure, red triangles are the experimental data, blue circles are the theoretical data, vertical black dashed lines guide the eye between the two datasets and dashed red lines are the prediction from the molecular constants presented in Table~\ref{tab:molecularConstants}.  In order to determine these points, we first converted the potential energy curves of Ref.~\cite{korek2009theoretical} from adiabatic to diabatic~\cite{facts1}, and then used LEVEL 8.0~\cite{le2016level} to calculate the positions of the $J^{\prime}=1$ eigenstate of the different vibrational levels.  The vibrational line spacings of the $A \ ^1\Sigma^+$ state that we observe show a smooth, gradual decrease for high vibrational states ($v^{\prime} > 20$), as expected.  For lower vibrational numbers, however, there are large shifts in several of the lines, and moderate shifts in others.  For example, the energy of $v^{\prime}=2$ line is shifted up by $\sim$10 cm$^{-1}$ above the general trend; in Fig.~\ref{fig:vibSpacing}(a) this appears as $T_{v^\prime=2}-T_{v^\prime=1}$ is too large by nearly 10 cm$^{-1}$ while $T_{v^\prime=3}-T_{v^\prime=2}$ is too small by 10 cm$^{-1}$.  Shifts of nearby vibrational lines, $v^\prime=10$ and 12, of the $b \ ^3\Pi_{0^+}$ state in the opposite direction suggest strong mixing between the $A \ ^1\Sigma^+$ and $b \: ^3\Pi_{0^+}$ state through spin-orbit interactions.  These perturbations recur over a rather large range of levels due to near coincidence between three times the vibrational spacing of the $A \ ^1\Sigma^+$ state with two times the vibrational spacing of the $b \: ^3\Pi_{0^+}$ state.  We will discuss these perturbations in more detail in the next section.  If we remove the most strongly perturbed vibrational states, the predicted vibrational spacings from the \textit{ab initio} calculations are in good agreement with the remaining observed spacings.  This is similar to our experience with other states in previous studies, such as the $d \ ^3\Pi$ state~\cite{stevenson2016d}, in which we found good general agreement between measured and calculated energy spacings between vibrational levels, but not with the well depth.  This disagreement manifests itself through two fewer vibrational levels than predicted for the $A \ ^1\Sigma^+ \ v^{\prime}$ series.  In Table~\ref{tab:molecularConstants}, we present a direct comparison of the derived molecular constants from theory and our data, confirming our earlier assertion of excellent agreement.

\section{${\bf A \ ^1\Sigma^+}$ - ${\bf b \ ^3\Pi_{0^+}}$ mixing} \label{sec:Abcomplex}

\begin{figure}[b]
	\includegraphics[width=8.6cm]{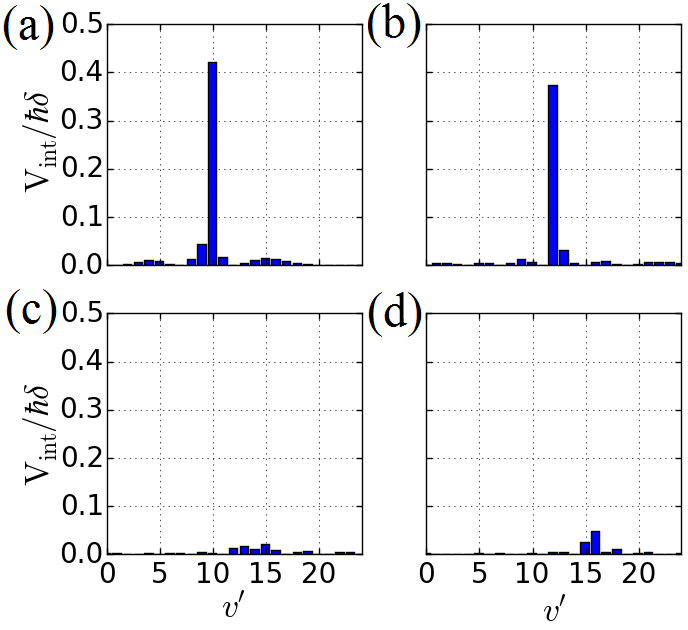}\\
	\caption{Calculated V$_{\rm int}=A_{Rb} |\langle \Psi_{b \ ^3\Pi} | \Psi_{A \ ^1\Sigma^+} \rangle|^2$/2, divided by the state energy difference, for $A \ ^1\Sigma^+$ (a) $v^\prime=2$, (b) $v^\prime=5$, (c) $v^\prime=8$ and (d) $v^\prime=11$ interacting with the vibrational levels of the $b \ ^3\Pi_{0^+}$ state.  The PECs are from Ref.~\cite{korek2009theoretical}, and we calculated the vibrational wavefunctions and FCFs using LEVEL 8.0~\cite{le2016level}.  $A_{Rb}=250$ cm$^{-1}$ is the atomic rubidium spin-orbit constant.}
	\label{fig:stateMixingBig}
\end{figure}

\begin{table*}[t]
	\centering
	\begin{tabular}{ccccccccccc}
		\hline \hline
		$A \ ^1\Sigma^+ \ v^{\prime}$ & E$- T_v$ (cm$^{-1}$) & $b \ ^3\Pi_{0^+} \ v^{\prime}$ & E$- T_v$ (cm$^{-1}$) & $\hbar \delta$ (cm$^{-1}$) & $(E_+ - E_-)$ (cm$^{-1}$) & $|c|$ & $|d|$ & $V_{\rm int}$ (cm$^{-1}$) & FCF & $\frac{V_{\rm int}}{FCF}$\\ \midrule[1.5pt]
		2 & 10.5 & 10 & -7.6 & 7.7 & 25.8 & 0.59 & 0.81 & 12.3 & 0.06 & 205\\
		5 & -4.5 & 12 & 6.1 & 0.7 & 11.3 & 0.68 & 0.73 & 5.6 & 0.03 & 187\\ \hline \hline
	\end{tabular}
	\caption{Parameters of the mixed states.  $\delta$ is the unshifted (bare) energy difference of the states, while $E_+ - E_-$ is the energy difference between the perturbed states.  $c$ and $d$ are the admixture coefficients determined through Eq.~(\ref{eq:c}) and (\ref{eq:d}).  $V_{\rm int}$/FCF as determined through this analysis is in reasonable agreement with $A_{Rb}/2 = 125$ cm$^{-1}$, where $A_{Rb}$ is the spin-orbit interaction strength in atomic rubidium.}
	\label{tab:fixingAssignments}
\end{table*}

In this section, we analyze state mixing between several vibrational levels of the $A \ ^1\Sigma^+$ and the $b \ ^3\Pi_{0^+}$ states.  The two largest perturbations seen in Fig.~\ref{fig:vibSpacing}(a) are due to mixing between the $A \ ^1\Sigma^+ \ v^{\prime}=2$ level with $b \ ^3\Pi_{0^+} \ v^{\prime}=10$, and between the $A \ ^1\Sigma^+ \ v^{\prime}=5$ level with $b \ ^3\Pi_{0^+} \ v^{\prime}=12$.   The mixing between states is proportional to~\cite{herzberg1951molecular} 
\begin{equation}
\frac{|\langle \Psi_{b \ ^3\Pi} | H_{\rm SO} | \Psi_{A \ ^1\Sigma^+} \rangle|^2}{\hbar \delta},
\label{eq:mixing}
\end{equation}
where $H_{\rm SO}$ is the Hamiltonian for the spin-orbit interaction, and $\hbar \delta$ is the energy difference between the unperturbed states.  
The vibrational factor of the wavefunctions $| \Psi \rangle $ in this expression implies that strong state mixing requires large Franck Condon overlap between two states, while the energy denominator requires small energy difference.  In Fig.~\ref{fig:stateMixingBig}, we show a rough estimate for the interaction strength caused by the spin-orbit effect divided by the state energy difference between the $A \ ^1\Sigma^+ \ v^{\prime}=2,$ 5, 8 and 11 states with the various vibrational levels $v^{\prime}$ of the $b \ ^3\Pi_{0^+}$ state.  We estimate the interaction strength with~\cite{weidemullerReview} $V_{\rm int}=A_{Rb} |\langle \Psi_{b \ ^3\Pi} | \Psi_{A \ ^1\Sigma^+} \rangle|^2$/2 where $A_{Rb}$ is the spin-orbit interaction in atomic rubidium, 250 cm$^{-1}$; we have found that this approximation roughly holds in the past~\cite{stevenson2016d}.  For $A \ ^1\Sigma^+ \ v^{\prime}=2$ and 5, mixing with one state is much stronger than any of the rest, justifying a two-state mixing model.  The $A \ ^1\Sigma^+ \ v^{\prime}=8$ state is hardly perturbed because it has a very small Franck Condon overlap with $b \ ^3\Pi_{0^+} \ v^\prime=14$ despite having nearly identical energies.  This is consistent with the story told by Fig.~\ref{fig:vibSpacing}, as neither $A \ ^1\Sigma^+ \ v^{\prime}=8$ nor $b \ ^3\Pi_{0^+} \ v^\prime=14$ appear significantly perturbed.  Finally, Fig.~\ref{fig:vibSpacing}(b) shows that $b \ ^3\Pi_{0^+} \ v^\prime=16$ is slightly perturbed, by $A \ ^1\Sigma^+ \ v^\prime=10$ and 11, which matches the small interaction strength shown in Fig.~\ref{fig:stateMixingBig}(d).  

For the $A \ ^1\Sigma^+ \ v^{\prime}=2$ and 5, which we approximate as a simple two-state mixing model, we calculated the admixture coefficients.  Using the treatment of mixed states in Ref.~\cite{herzberg1951molecular}, we write the mixed states as
\begin{equation}
| \Psi_- \rangle = c | \Psi_{A \ ^1\Sigma^+} \rangle - d | \Psi_{b \ ^3\Pi} \rangle 
\end{equation}
and
\begin{equation}
 | \Psi_+ \rangle = d | \Psi_{A \ ^1\Sigma^+} \rangle + c | \Psi_{b \ ^3\Pi} \rangle,
\label{eq:mixedStates}
\end{equation}
where $| \Psi_{A \ ^1\Sigma^+} \rangle$ and $| \Psi_{b \ ^3\Pi} \rangle$ are the bare states.  The energies of states $| \Psi_- \rangle $ and $| \Psi_+ \rangle $ are $E_-$ and $E_+$, respectively.  As we have done in the past~\cite{stevenson2016direct}, we can use the expected energy and observed energy of mixed states to estimate their mixing; we use the molecular constants of Table~\ref{tab:molecularConstants} to predict where the unperturbed state should lie.  After we have a prediction for the unperturbed state location we can solve for the admixture coefficients
\begin{equation}
c^2 = \frac{1}{2} \left[ 1 - \frac{\hbar \delta}{E_+ - E_-}  \right] 
\label{eq:c}
\end{equation}
and
\begin{equation}
d^2 = \frac{1}{2} \left[ 1 + \frac{\hbar \delta}{E_+ - E_-}  \right].
\label{eq:d}
\end{equation}
Additionally, we calculate the interaction strength, V$_{\rm int}$, using V$_{\rm int} = \frac{1}{2} \sqrt{(E_+ - E_-)^2- (\hbar \delta )^2}$ from Ref.~\cite{stevenson2016direct}.  All of the relevant parameters for the strongest mixed $A \ ^1\Sigma^+ - b \ ^3\Pi_{0^+}$ states that we observe and the derived admixture coefficients and interaction strengths are presented in Table~\ref{tab:fixingAssignments}.  For both of these states, $V_{\rm int}/$FCF is within a factor of two of $A_{Rb}/2$, showing approximate agreement with the model of Ref.~\cite{weidemullerReview}. These mixed states are critical for some future experiments, as discussed in the next section.

\section{Outlook}\label{sec:Outlook}

We performed spectroscopy on three electronic states in LiRb: $C \ ^1\Sigma^+$, $A \ ^1\Sigma^+$, and $b \ ^3\Pi_{0^+}$, with the intention of using the data presented here to guide future work.  

1) In future work, we plan to explore the $C \ ^1\Sigma^+ \ v^{\prime}=22$ as an intermediate state for STIRAP to transfer population from $v^{\prime \prime}=43$ to $v^{\prime \prime}=0$.  PA to the $2(1)-4(1)$ mixed state produces $v^{\prime \prime}=43$ molecules at a rate of $\approx 3\times10^5$ molecules/second.  For future applications, which we want to transfer these molecules to $v^{\prime \prime}=0$.  We observed vibrational levels $v^{\prime}=7$, 9, 12, 13 and 26-45, and saw very regular vibrational spacing.  This is reflected in our harmonic constant, $\omega_e$, which has an uncertainty of only 0.9 cm$^{-1}$.  Such uncertainty combined with depletion spectra should allow us to find $C \ ^1\Sigma^+ \ v^{\prime}=22$ easily.

2) We are interested in studying $b \ ^3\Pi_{0^+} \ v^{\prime} = 0 - 2$; as suggested by Ref.~\cite{you2016ab}, these states could be used to laser-cool ground state LiRb molecules.  In the present experiment, we exploited the strong mixing in the $A \ ^1\Sigma^+ - b \ ^3\Pi_{0^+}$ complex and we were able to observe $b \ ^3\Pi_{0^+} \ v^{\prime} = 8 - 18$.  Unfortunately, finding more deeply bound vibrational levels will be challenging.  We noticed that the amount of mixing between $A \ ^1\Sigma^+ - b \ ^3\Pi_{0^+}$ states dictates the strength of the $b \ ^3\Pi_{0^+}$ state depletion resonance, and we expect the mixing to decrease as the states get more deeply bound.  Further, the $b \ ^3\Pi_{0^+} \ v^{\prime} = 8 \leftarrow v^{\prime \prime}=2$ transition had a linewidth of only 1 GHz at full depletion power (compare with $A \ ^1\Sigma^+ \ v^{\prime} = 0 \leftarrow v^{\prime \prime}=2$ transition which was well over 100 GHz in linewidth), which makes this a daunting search.  However, we expect our molecular constants and those of Refs~\cite{you2016ab} and \cite{korek2009theoretical} to be a helpful guides.

3) Our data on mixing between $A \ ^1\Sigma^+ \ v^\prime=5$ and $b \ ^3\Pi_{0^+} \ v^{\prime}=12$ suggests that this state pair is at a nearly 50-50 admixture.  This pair of states combine many features that lead us to believe they may be useful in direct short range PA.  The $A \ ^1\Sigma^+ \ v^\prime=5$ state has large FCF overlap with the ro-vibronic ground state, and the inner turning point of $b \ ^3\Pi_{0^+} \ v^{\prime}=12$ nearly lines up with the outer turning point of the triplet scattering state.  This is a good indicator of states that have potential for use in short range PA~\cite{blasingsce16}.  Finally, the wavelength for PA to this state is 1562 nm which can be accessed with commercial fiber amplifiers.  This is important because we will have watts of PA power available which will help overcome the weak PA rates common to most short range PA experiments.  In total, $A \ ^1\Sigma^+ \ v^\prime=5$ and $b \ ^3\Pi_{0^+} \ v^{\prime}=12$ may be nearly ideal candidates for producing ro-vibronic ground state molecules by short range PA.  As a rough estimate this could produce up to $5 \times 10^4$ ro-vibronic ground state molecules/second~\cite{facts2}.

\section{Conclusion}

In conclusion, we used molecules produced by PA to a $2(1) - 4(1)$ state to study interesting bound states in LiRb.  We used RE2PI to map out the top 20 vibrational levels and four lower vibrational levels of the $C \ ^1\Sigma^+$ state.  We used depletion spectroscopy to study the 30 most deeply bound $A \ ^1\Sigma^+$ vibrational levels and discovered that the $A \ ^1\Sigma^+ - b \ ^3\Pi_{0^+}$ mixing is quite strong in this region.  We capitalized on the $A \ ^1\Sigma^+ - b \ ^3\Pi_{0^+}$ mixing to study $b \ ^3\Pi_{0^+} \ v^{\prime} = 8 - 18$.  This data will guide our future work in both STIRAP and laser cooling LiRb molecules.

We are happy to acknowledge useful conversations with and help in calculating the short range PA rate from Jes\'{u}s P\'{e}rez-R\'{\i}os; advice on assembling the depletion laser provided by George Toh; conversations with Sourav Dutta that sent us down the $C \ ^1\Sigma^+$ spectroscopy path; and university support of this work through the Purdue OVPR AMO incentive grant.


\begin{thebibliography}{24}
\bibitem[{fac({\natexlab{a}})}]{facts1}
	\bibinfo{note}{Converting adiabatic PECs to diabatic PECs means that avoided
		crossings no longer avoid. Additionally, our conversion is not perfect and
		resulted in small perturbations to the theoretical vibrational series near $A
		\ ^1\Sigma^+ \ v^{\prime}=0$ and $b \ ^3\Pi_{0^+} \ v^\prime=8$}.
\bibitem[{fac({\natexlab{b}})}]{facts2}
	\bibinfo{note}{Our estimate is based on calculations of excited state Franck -
		Condon overlap with the triplet scattering state done by Jes\'{u}s
		P\'{e}rez-R\'{\i}os. We used this to determine how strong PA to $A \
		^1\Sigma^+ \ v^{\prime}=5$ is relative to $d \ ^3\Pi_{0^-} \ v^{\prime}=0$,
		measured in Ref~\cite{blasingsce16} to be $5 \times 10^3$ molecules/second.
		Finally, we multiplied this PA rate by the branching ratio to $v^{\prime
			\prime}=0$ and augmented it by the power increase we will get from having
		much more PA power available. Our estimate is made assuming 1 W of PA power
		and is by no means an upper limit on the production rate.}
\end{thebibliography}
\end{document}